\documentclass[pra,twocolumn,superscriptaddress]{revtex4-1}
\usepackage{amsmath}
\usepackage{amssymb}
\usepackage{graphicx}
\usepackage{color}

\begin{document}

\title{Tachyonic Dispersion in Coherent Networks}

\author{Y.~D.~Chong}
\email{yidong@ntu.edu.sg}

\affiliation{Division of Physics and Applied Physics, School of Physical and Mathematical Sciences, Nanyang Technological University, Singapore 637371, Singapore}

\affiliation{Centre for Disruptive Photonic Technologies, Nanyang Technological University, Singapore 637371, Singapore}

\author{M.~C.~Rechtsman}

\affiliation{Department of Physics, The Pennsylvania State University, University Park, PA 16802, USA}

\begin{abstract}
We propose a technique to realize a tachyonic band structure in a
coherent network, such as an array of coupled ring resonators.  This
is achieved by adding ``PT symmetric'' spatially-balanced gain and
loss to each node of the network.  In a square-lattice network, the
quasi-energy bandstructure exhibits a tachyonic dispersion relation,
centered at either the center or corner of the Brillouin zone.  There
is one tachyonic hyperboloid in each gap, unlike in PT-symmetric
tight-binding honeycomb lattices where the hyperboloids occur in
pairs.  The dispersion relation can be probed by measuring the peaks
in transmission across a finite network as the gain/loss parameter is
varied.
\end{abstract}

\maketitle

Most wave theories, including but not limited to quantum mechanics and
classical electromagnetism, are formulated using equations of motion
with Hermitian Hamiltonians.  In quantum mechanics, Hermiticity
ensures the general conservation of total probability under time
evolution; conversely, its violation describes amplification (gain)
and/or loss.  Thus, for instance, non-Hermitian Hamiltonians are used
in effective theories of decaying quantum systems, in which the
wavefunction can leak away into unmonitored degrees of freedom.  In
optical physics, gain and loss processes are even more ubiquitous, in
the context of the emission and absorption of light, and the
prescriptions for dealing with these processes (e.g., introducing
complex frequency-domain dielectic permittivities) are similarly well
known.

Several years ago, Bender and co-workers made the striking observation
that in systems possessing parity-time ($PT$) symmetry, corresponding
to spatially-balanced gain and loss, the Hamiltonian can have purely
real eigenvalues (i.e., probability-conserving eigenstates) despite
being non-Hermitian \cite{Bender,Bender02}.  Subsequently, a series of works showed both theoretically and experimentally that this
effect could be demonstrated in optical structures, using optical gain
and loss
\cite{ElGanainy,Kostas1,Kostas2,Musslimani,Guo,Ruter,Peschel}.  In
$PT$ symmetric optical lattices \cite{Kostas2, Peschel}, the photonic
band structure has quite unusual features: the band energies can be
real in one region of the Brillouin zone, where the Bloch eigenstates
are $PT$ symmetric, and complex in another region where the $PT$
symmetry is spontaneously broken.

For 2D lattices, Szameit \textit{et al.}~showed that the $PT$
symmetry-breaking phenomenon has a startling interpretation in terms
of emergent ``tachyons'': hypothetical superluminal particles which
are not known to exist in nature \cite{tachyons}.  A two-dimensional
honeycomb lattice can be realized using an array of coupled optical
waveguides.  In the Hermitian case, the bandstructure is
graphene-like, featuring a pair of linear band crossing points
(``Dirac points'') with band velocity $v_D$ \cite{Peleg,RechtsmanFTI}.
When gain and loss are added to alternating sites of the honeycomb
lattice, the Bloch Hamiltonian becomes non-Hermitian, and in the
vicinity of each Dirac point it takes the form of a Dirac Hamiltonian
with imaginary mass.  The eigenstates are tachyons whose group
velocities are larger than $v_D$.  In fact, the group velocities
become infinite along a ``critical'' ring in $k$-space surrounding
each Dirac point, corresponding to the $PT$ symmetry breaking
transition points of the Bloch Hamiltonian.  A similar dispersion
relation has also recently been realized in Ref.~\cite{Zhen}. However,
it should be noted that the notion of group velocity as the slope of
the dispersion relation must be reevaluated in non-Hermitian systems.
Using the Hellman-Feynman theorem, it has been shown that as the
critical ring is reached in k-space, significant corrections to this
definition of the group velocity arise
\cite{superluminal,SchomerusGroupVelocity,Kostas3}.

This paper describes an alternative way to realize a tachyon-like
bandstructure, using a 2D network of coherent waves
\cite{ChalkerCo,kramersurvey} with non-unitary evolution.  A ``network
model'', unlike the tight-binding models commonly used in
condensed-matter and optical physics, does not describe a lattice in
terms of a Hamiltonian.  Instead, it uses an evolution matrix to
describe the propagation of waves through a network of directed links
and nodes.  As discussed below, such networks can be realized in a
variety of ways, such as coupled optical resonator lattices
\cite{Shayan, hafezi2,Mercedes, Zhang}, microwave networks
\cite{Hu,Gao}, and RF circuits \cite{Ningyuan}.  Network models can
produce bandstructures with various unusual features that are not
found in static Hamiltonian models \cite{Pasek}; in fact they can be
mapped to the class of ``Floquet'' systems, described by Hamiltonians
that vary periodically in time
\cite{Lindner,Cayssol,Levin,RechtsmanFTI}.  As we shall see,
introducing $PT$ symmetric gain and loss to a \textit{square}-lattice
(not honeycomb) network yields a bandstructure with tachyonic Dirac
dispersion relations.  But unlike the previously-studied tight-binding
honeycomb lattice, where the tachyonic Dirac hyperboloids occur in
pairs, this network bandstructure contains a single hyperboloid in
each gap.  Finally, we will show how the tachyonic dispersion
relation's critical $k$-vector can be measured through transmission
experiments across finite networks.

\begin{figure}
  \centering\includegraphics[width=0.48\textwidth]{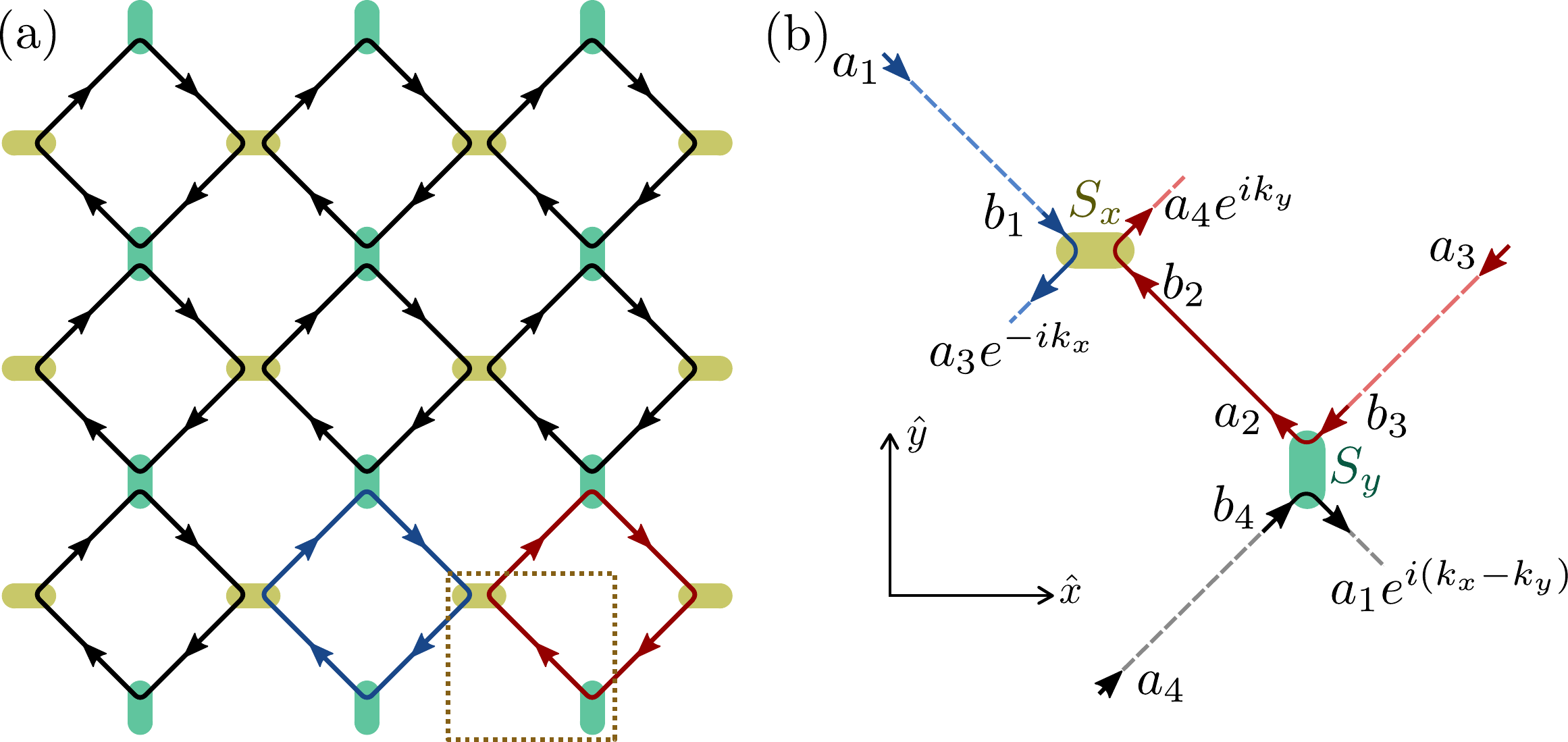}
  \caption{(a) Schematic of a chiral square-lattice network.  Waves
    propagate directionally in the links (arrows), forming clockwise
    loops, and couple at the nodes where adjacent loops approach each
    other (shaded ovals).  Two of the loops are highlighted in red and
    blue for ease of comparison with (b) and (c).  (b) Detail of a
    unit cell of the network, showing the wave amplitudes entering and
    leaving each node.  These amplitudes enter into
    Eq.~(\ref{couplings}).}
  \label{fig:schematic}
\end{figure}

Consider the network model shown schematically in
Fig.~\ref{fig:schematic}.  It consists of links and nodes, where each
link carries a one-directional wave described by a complex scalar
amplitude; these are arranged in a 2D square lattice, with each cell
containing four links arranged in a chiral loop \cite{ChalkerCo}.
Adjacent loops are coupled at the nodes, which are described by
$2\times2$ scattering matrices.  This model was first introduced for
studying the transport properties of disordered quantum Hall systems
\cite{ChalkerCo}; it captures the essential features of a disordered
2D electron gas in strong magnetic fields, where the electronic orbits
follow chiral ``race-tracks'' along the equipotentials of a disordered
potential landscape, and can tunnel to adjacent race-tracks at
potential saddle-points \cite{ChalkerCo}.  There is now an extensive
literature on the use of network models for studying electronic
transport; see Ref.~\cite{kramersurvey} for a survey.

Recently, researchers have implemented chiral networks in classical
electromagnetic settings.  One type of realization is an on-chip
coupled resonator lattice \cite{hafezi,hafezi2}, of the sort proposed
and experimentally studied by Hafezi \textit{et al.}  Optical ring
resonators are arranged in a lattice, playing the role of the
network's chiral loops.  Each pair of adjacent resonators is coupled
by an auxiliary ring waveguide, which acts as a node.  Due to local
momentum conservation at the inter-waveguide interfaces, the optical
modes of the lattice decouple into one set of modes where light
propagates clockwise in the main rings, and another that is
counter-clockwise; each set maps onto a network model.  Such resonator
lattices can exhibit topological edge states and fractal Hoftadter
spectra \cite{hafezi,hafezi2}, as well as topological transitions and
anomalous topological phases \cite{Liang,Liang2,Pasek,Gao}.  A chiral
network can also be realized using a microwave circuit \cite{Hu}.  The
nodes of the network are implemented using directional couplers;
auxiliary rings are not necessary, since the microwave components need
not be strictly planar.  The chirality of the network can be enforced
using microwave isolators.


Regardless of the network model's underlying implementation, its
properties can be described theoretically in terms of evolution
matrices.  And for a disorder-free, spatially infinite network, the
evolution matrix description gives rise to a bandstructure
\cite{HoChalker,Pasek}.  We briefly review the procedure.
Fig.~\ref{fig:schematic}(b) shows a schematic of one cell of the
network, which contains two couplers connecting adjacent loops along
the $\hat{x}$ and $\hat{y}$ directions.  These couplers can be
described by scattering matrices $S_x$ and $S_y$.  We denote the four
input wave amplitudes into these couplers by $\{b_1, \dots, b_4\}$,
and the wave amplitudes on the other side of those links by
$\{a_1,\dots,a_4\}$.  We assume each link has equal phase delay
$\phi$, so that $b_n = e^{i\phi} a_n$.  For the moment, we ignore gain
and loss, so that $S_x$ and $S_y$ are unitary and $\phi$ is real.
Using Bloch's theorem, we can relate the input and output amplitudes
as follows:
\begin{align}
  \begin{aligned}
  S_x \begin{bmatrix}b_1\\b_2
  \end{bmatrix} &= \begin{bmatrix}a_3 e^{-ik_x} \\ a_4 e^{ik_y}
  \end{bmatrix} \\
  S_y \begin{bmatrix}b_3\\b_4
  \end{bmatrix} &= \begin{bmatrix}a_2 \\ a_1 e^{i(k_x-k_y)}
  \end{bmatrix},
  \end{aligned}
  \label{couplings}
\end{align}
where $k = [k_x,k_y]$ is the Bloch wavevector, with the lattice
spacings normalized to 1.  These can be combined into a single
$4\times4$ eigenvalue equation:
\begin{equation}
  U(k)  \begin{bmatrix}b_1\\b_2\\b_3 \\ b_4
  \end{bmatrix} = e^{-i\phi} \begin{bmatrix}b_1\\b_2\\b_3 \\ b_4
  \end{bmatrix},
  \label{Uk}
\end{equation}
where
\begin{align}
  U(k) &= \begin{bmatrix}
    \mathbf{0} & W_y(k) S_y \\ W_x(k) S_x & \mathbf{0}
  \end{bmatrix} \\
  W_x(k) &= \begin{bmatrix}e^{ik_x} & 0 \\ 0 & e^{-ik_y}
    \end{bmatrix} \label{Wx} \\
  W_y(k) &= \begin{bmatrix}0 & e^{-i(k_x-k_y)} \\ 1 & 0
  \end{bmatrix}.
  \label{Wy}
\end{align}
For each $k$, $U(k)$ is unitary, and its eigenvalues are
$\exp(-i\phi)$, where the ``quasi-energies'' $\phi$ are the discrete
values of the link delay for which modes can propagate with the given
$k$.  Because the bandstructure is defined by an evolution operator,
$U(k)$, rather than a Hamiltonian, it falls into the same class as the
``Floquet'' bandstructures describing periodically-driven lattices
\cite{Lindner,Cayssol,Levin,RechtsmanFTI}.  Note that $\phi$ is an
angle variable, unlike the energy occurring in a conventional
Hamiltonian eigenproblem.

\begin{figure}
  \centering\includegraphics[width=0.4\textwidth]{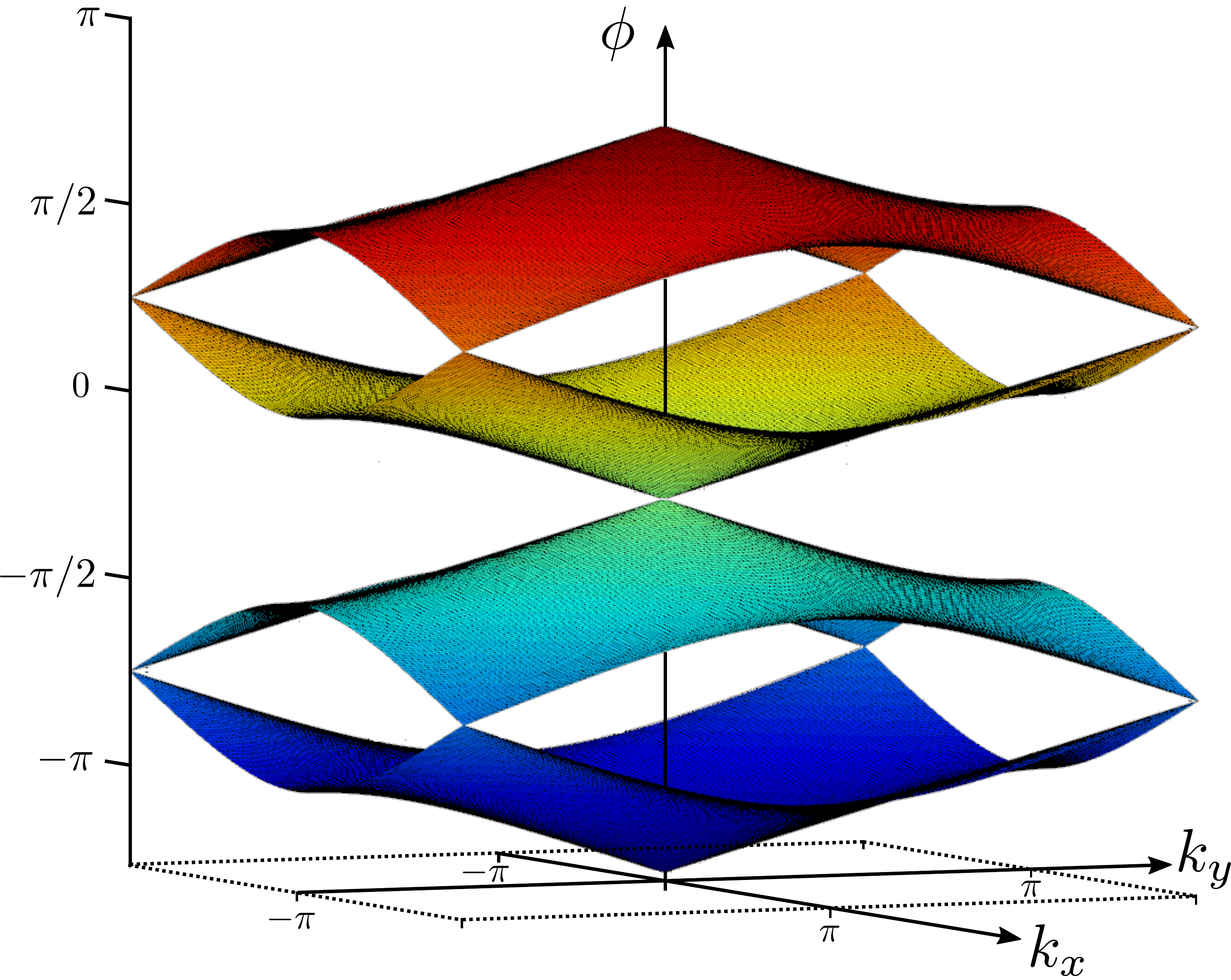}
  \caption{Bandstructure of the network model for $\theta_x = \theta_y
    = \pi/4$.  There are Dirac points at $\phi = \{-\pi/4, 3\pi/4\}$
    and $k_x = k_y = 0$, and at $\phi = \{-3\pi/4, \pi/4\}$ and $k_x =
    k_y = \pm \pi$.  }
  \label{fig:bandstructure}
\end{figure}

The bandstructure depends on the choice of the $2\times2$ unitary
matrices $S_x$ and $S_y$, each parameterized by four Euler angles.  It
turns out that the bandstructure topology is determined only by one
pair of Euler angles, denoted by $\theta_{x,y}$ in Ref.~\cite{Liang},
which parameterize the coupling strengths between adjacent loops.  The
gaps in the quasi-energy bandstructure close when $\theta_x + \theta_y
= \pi/2$ \cite{Liang}.  For simplicity, we fix the other Euler angles
so that
\begin{equation}
  S_{\mu} = \begin{bmatrix} \cos\theta_\mu & i\sin\theta_\mu
    \\ i\sin\theta_\mu & \cos\theta_\mu \end{bmatrix}, \;\;\;\mathrm{for}
  \;\;\; \mu \in \{x,y\}.
  \label{Smu}
\end{equation}
This describes a coupler which is symmetric under $180^\circ$
rotations [see Fig.~\ref{fig:schematic}(b)], and behaves the same when
the order of the two inputs, and the two outputs, are simultaneously
swapped.

Fig.~\ref{fig:bandstructure} shows the spectrum for $\theta_x =
\theta_y = \pi/4$.  There are four bands, joined by Dirac points at
\begin{align}
  \begin{aligned}
    &\left\{ \begin{array}{l} k_x = k_y = 0, \\
      \phi \in \{-\pi/4, 3\pi/4\} 
    \end{array}
    \right. \\
    &\left\{ \begin{array}{l} k_x = k_y = \pi, \\
      \phi \in \{-3\pi/4, -\pi/4\}.
    \end{array}
    \right.
    \label{Dirac points}
  \end{aligned}
\end{align}
These Dirac points can be conveniently derived by taking the squared
matrix \cite{HoChalker}:
\begin{align}
  \begin{aligned}
  U^2(k) &= \begin{bmatrix}\mathcal{U}_1(k) & \mathbf{0} \\ \mathbf{0} &
    \mathcal{U}_2(k)
  \end{bmatrix}, \\
  \mathcal{U}_1 &= W_y S_y W_x S_x, \;\;\;
  \mathcal{U}_2 = W_x S_x W_y S_y.
  \end{aligned}
  \label{U2}
\end{align}
$U^2$ consists of two blocks with identical spectra, and its
eigenvalues are $\exp(-2i\phi + 2m\pi)$ for $m \in \mathbb{Z}$.
Focusing on the first block, we let $\theta_\mu = \pi/4 + \Delta
\theta$, and expand to first order in $k_\mu$ and $\Delta \theta$.
The result is
\begin{equation}
  \mathcal{U}_1(k) \approx \exp\left\{-i\left[H_D(k) -
    \frac{\pi}{2}\right] + \cdots \right\},
\end{equation}
where
\begin{equation}
 H_D(k) \equiv \frac{\sigma_y + \sigma_z}{2} \, k_x + \frac{\sigma_y -
   \sigma_z}{2} \, k_y - 2\Delta\theta\, \sigma_x.
  \label{Dirach}
\end{equation}
This is a Dirac Hamiltonian with mass $2\Delta\theta$ and band
velocity $v_D = 2^{-3/2}$.  Another pair of Dirac points is obtained
by expanding $k_\mu = \pi + \kappa_\mu$.  The result is
\begin{equation}
  \mathcal{U}_1 \approx \exp\left\{-i\left[H_D(\kappa) +
    \frac{\pi}{2}\right] + \cdots \right\},
\end{equation}
which gives the second set of Dirac points in Eq.~(\ref{Dirac
  points}).

In the critical bandstructure ($\Delta\theta = 0$), there is a single
Dirac cone in each gap.  This contrasts with the more familiar case of
time-reversal symmetric honeycomb lattices (e.g.~graphene), where the
Dirac points occur in pairs, which is a manifestation of the ``fermion
doubling'' principle \cite{haldane88}.  The unpaired Dirac point in
the network model's bandstructure is reminiscent of the unpaired
massless chiral relativistic fermions which occur at the Haldane
model's critical points under broken inversion and time-reversal
symmetry \cite{haldane88}.  In the network model, those symmetries are
likewise explicitly broken.

Interestingly, this critical bandstructure cannot be described
adequately by an effective Hamiltonian $H_{\mathrm{eff}}(k) =
i\log[U(k)]$.  This is because the quasi-energy bandstructure is
completely ungapped; the eigenvalues of $U(k)$, over all $k$, cover
the unit circle.  Thus, there is no way to assign the logarithm's
branch cut, without there being a locus of $k$-points where some
eigenvalues of $U(k)$ cross the cut.  There is no
$H_{\mathrm{eff}}(k)$ which can be ``smoothly'' defined over all $k$
in the Brillouin zone.

If we tune the coupling strengths away from the critical point, the
quasi-energy bandstructure becomes gapped.  For $\Delta\theta < 0$, it
is topologically trivial, and for $\Delta\theta > 0$ it is
topologically non-trivial \cite{Liang,Pasek}.  In both cases, however,
it can be shown that every band has zero Chern number \cite{Pasek}.
This happens because, in the critical bandstructure, each band had a
Dirac point \textit{above} and \textit{below}, a situation that is
possible because $\phi$ is an angular variable and hence not bounded
above or below.  Hence, for $\Delta\theta > 0$ the network is in an
``anomalous Floquet insulator'' phase, exhibiting topological edge
states despite all bands having zero Chern numbers.  Similar anomalous
phases are also known to occur in periodically-driven Floquet
topological insulators \cite{Lindner,Cayssol,Levin}.

\begin{figure}
  \centering\includegraphics[width=0.4\textwidth]{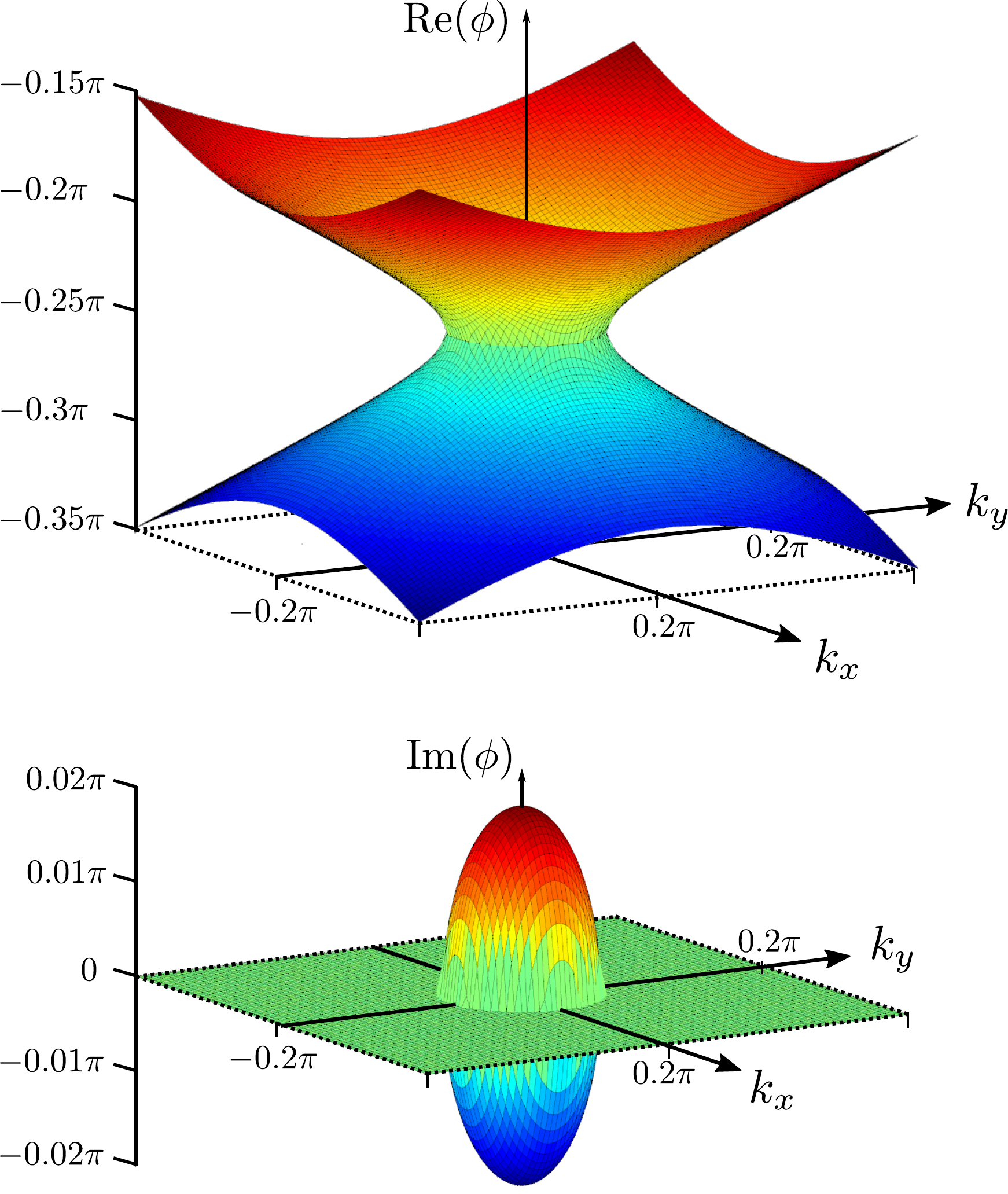}
  \caption{Section of the network bandstructure for $\theta_x =
    \theta_y = (0.25 + 0.02i)\pi$.  Each of the Dirac points becomes a
    hyperboloid.  We have zoomed in on the hyperboloid at $\phi \sim
    -\pi/4$, $k_x \sim k_y \sim 0$.  }
  \label{fig:tachyon}
\end{figure}

We are now ready to consider a network containing gain and loss.  In
the honeycomb lattice, Szameit \textit{et al.}~have previously shown
that adding $PT$ symmetric gain and loss to the alternate sublattices
distorts the bandstructure's Dirac cones into hyperboloids,
corresponding to two species of emergent tachyons \cite{tachyons}.
This is caused by the Bloch states near each Dirac point undergoing
spontaneous $PT$ symmetry breaking.  In the $PT$ symmetric region, the
bands are real and have group velocity exceeding the Dirac velocity
$v_D$.  The group velocity approaches infinity at the waists of the
hyperboloids, which are the $PT$ symmetry breaking points of the Bloch
Hamiltonian.

In the network model, tachyonic behavior can arise by setting
$\Delta\theta$ to be imaginary.  This can potentially be realized in
the optical resonator domain by using `auxiliary rings' that lie in
between the principal rings, and are optically pumped and thus have
gain (unpumped rings would naturally exhibit loss).  A candidate
platform would be that used in Ref.~\cite{Mercedes}.  In the context
of microwave networks, auxiliary directional couplers could be used in
a similar way, in combination with amplifiers.  According to
Eq.~(\ref{Dirach}), this gives the effective Dirac Hamiltonian an
imaginary mass.  For $\Delta\theta = i \gamma$, the coupling matrices
become
\begin{equation}
  S_{\mu} = \begin{bmatrix}
    \alpha & i\alpha^* \\ i\alpha^* & \alpha
  \end{bmatrix}, \;\;\;
  \alpha = \frac{\cosh\gamma-i\sinh\gamma}{\sqrt{2}}.
  \label{Smupt}
\end{equation}
This yields the bandstructure shown in Fig.~\ref{fig:tachyon}.  Each
Dirac cone becomes a hyperboloid, corresponding to a tachyonic
dispersion relation.  Since there was originally only one Dirac cone
per gap, the hyperboloids are unpaired, unlike in the $PT$ symmetric
honeycomb lattice \cite{tachyons}.  The band quasi-energies are all
real, except for the regions of $k$ inside the waists of the
hyperboloids.  Using Eq.~(\ref{Dirach}), we find the critical
wavenumbers
\begin{equation}
  k_c(\gamma) = 2^{3/2} \gamma + O(\gamma^2).
  \label{criticalk}
\end{equation}

The coupling matrix of Eq.~(\ref{Smupt}) has the same $180^\circ$
rotational symmetry as the previously-discussed unitary coupling
matrix of Eq.~(\ref{Smu}).  However, for $\gamma \ne 0$ it is
manifestly non-unitary.  This may be seen from the eigenvalues
$\sigma_\pm = \alpha \pm i \alpha^*$, whose magnitudes are
$|\sigma_\pm| = e^{\pm \gamma}$.  The corresponding eigenvectors are
$[1;\pm 1]$; one of these eigenvectors is amplified, and the other is
damped by an equal and opposite amount.  This is very similar to the
behavior of scattering matrices derived from the wave equation in $PT$
symmetric media \cite{PT0,PT1,PT2}.  Furthermore, the coupling matrix
can be decomposed as
\begin{equation}
  S_\mu = \mathcal{S}_0
\begin{bmatrix}\sigma_+ & 0 \\ 0 & \sigma_-
  \end{bmatrix} \mathcal{S}_0, \;\;
\mathrm{where}\;\; \mathcal{S}_0 \equiv
\begin{bmatrix}\frac{1}{\sqrt{2}} & \frac{1}{\sqrt{2}}
    \\ \frac{1}{\sqrt{2}} & - \frac{1}{\sqrt{2}} \end{bmatrix}.
\end{equation}
Thus, such a coupler can be implemented by passing the inputs through
a unitary 50:50 coupler described by $\mathcal{S}_0$, applying
balanced gain and loss to the results, and then re-mixing through a
second $\mathcal{S}_0$ coupler.


How might the tachyonic bandstructure be experimentally verified, whether in the context of optical ring resonators or microwave networks?  One
possibility is to construct a wavepacket and show that its group
velocity can exceed the effective Dirac velocity, as discussed in
Ref.~\cite{tachyons}.  For the network model, this approach could work
if $\phi$ is proportional to frequency, and the coupling parameters
are approximately frequency-independent \cite{Liang2}.  However, the
natural quantities to study in a network model are the steady-state
reflection and transmission for fixed $\phi$.  Here, we present an
alternative experimental approach for probing the tachyonic
bandstructure, based on measuring the transmission across a set of
finite networks.

\begin{figure}
  \centering\includegraphics[width=0.48\textwidth]{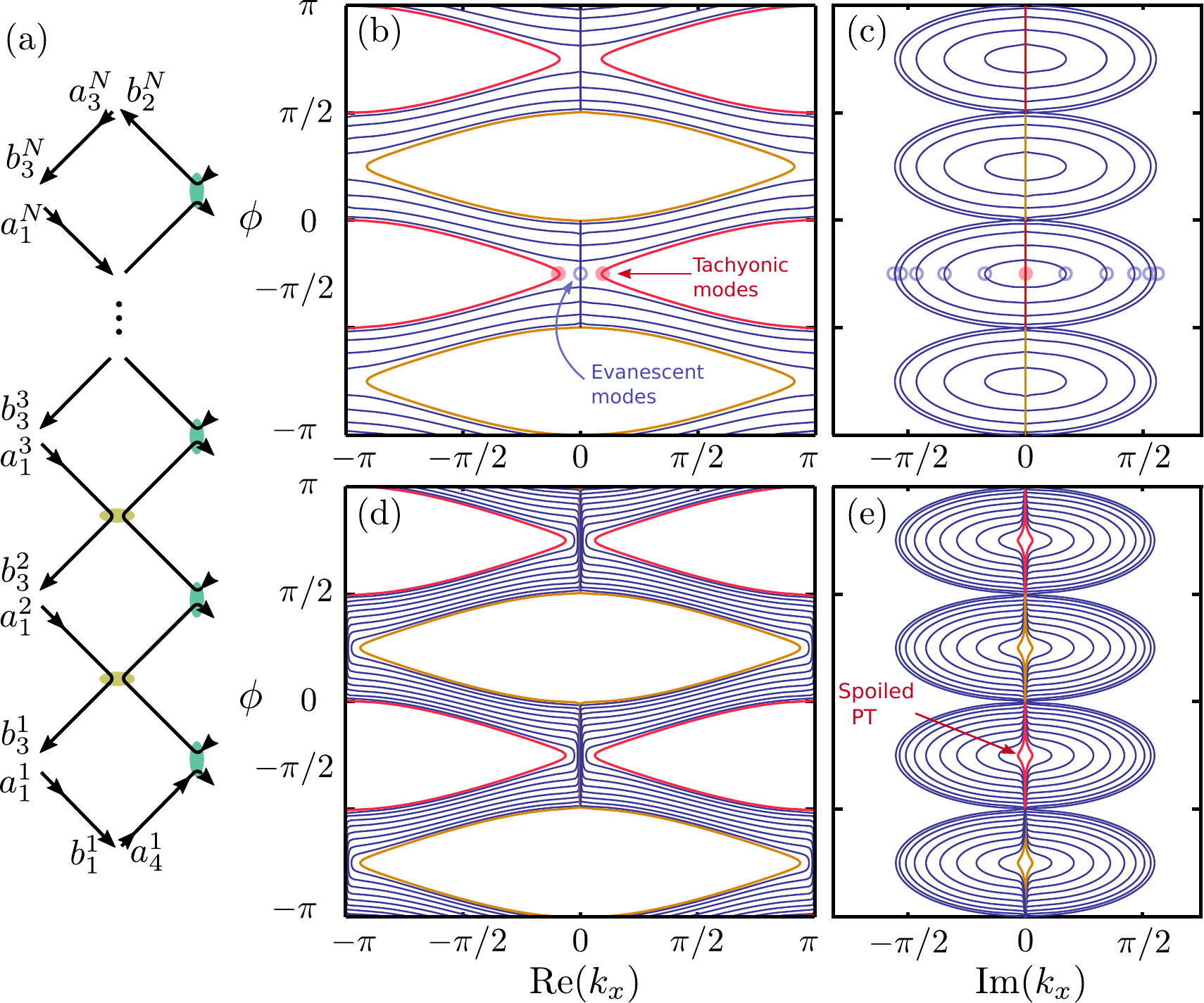}
  \caption{ (a) Schematic of a semi-infinite network ``strip'' of
    width $N$ in the $\hat{y}$ direction.  (b)--(c) Band diagram of
    real $\phi$ versus complex $k_x$, with periodic boundary
    conditions applied at the upper and lower edges of the strip.
    Here, we take $N = 10$ and coupling parameter $\theta = \pi/4 +
    0.1i$.  At $\phi = -\pi/4$, the red arrow indicates a pair of
    tachyon modes, which have $\mathrm{Re}[k_x] \ne 0$ (left plot) and
    $\mathrm{Im}[k_x] = 0$ (right plot).  At the same quasi-energy,
    there is a set of evanescent modes with $\mathrm{Re}[k_x] = 0$ and
    $\mathrm{Im}[k_x] \ne 0$.  (d)--(e) Band diagram with Dirichlet
    boundary conditions at the strip edges.  The Dirichlet boundary
    conditions spoil the PT symmetry, causing the tachyonic states to
    have small but non-zero $\mathrm{Im}[k_x]$.  }
  \label{fig:projected}
\end{figure}

Let us first examine the bandstructure of a ``strip'' of network,
shown schematically in Fig.~\ref{fig:projected}(a).  The strip extends
infinitely in the $\hat{x}$ direction, and has a width of $N$ cells in
the $\hat{y}$ direction.  There are two useful choices of boundary
conditions that we can impose on the edges of the strip.  Firstly, we
can impose periodic edges by making row $N+1$ equivalent to row $1$
(i.e., rolling the strip into the surface of cylinder).  Secondly, we
can impose ``Dirichlet'' edges by terminating the network at the edges
of the strip, setting $b_1^1 = a_4^1$ and $b_2^{N} = a_3^{N}$.  (We
could also introduce phase factors into these edge relations; but that
generates additional non-topological edge states, which we are not
interested in here.)

Fig.~\ref{fig:projected} shows the band diagram of real-$\phi$ versus
complex-$k_x$, for the semi-infinite strip.  This band diagram is
calculated from the eigenvalues of the transfer matrix across one
$\hat{x}$ period of the strip \cite{Pasek}.  The reason for plotting
real $\phi$ versus complex $k_x$, rather than complex $\phi$ versus
real $k_x$, is that we will be interested in the propagation of modes
at a fixed \textit{real} quasi-energy $\phi$, chosen to correspond to
one of the Dirac points.  We first focus on
Fig.~\ref{fig:projected}(b)--(c), which shows the case of periodic
edges.  For each real $\phi$, all the modes are either purely
propagating (real $k_x$), or purely evanescent (imaginary $k_x$).  The
evanescent modes are completely non-propagating ($\mathrm{Re}[k_x] =
0$), and are similar to the evanescent modes which occur within the
band gaps of ordinary Hermitian systems.  As for the propagating
modes, there are specific branches of these modes which have tachyonic
dispersion relations, and are highlighted in red and gold in the
figure.  As indicated in Fig.~\ref{fig:projected}(c), these modes
propagate with no amplification nor dissipation ($\mathrm{Im}[k_x] =
0$).

Fig.~\ref{fig:projected}(d)--(e) shows the band diagram for Dirichlet
edges.  In this case, the modes are no longer purely propagating or
purely evanescent, but have complex $k_x$.  This happens because the
edge conditions spoil the PT symmetry of the network.  Nonetheless,
the projected band diagram remains qualitatively similar to
Fig.~\ref{fig:projected}(b)--(c).  In particular, there are tachyon
modes which are weakly damped (small $|\mathrm{Im}(k_x)|$) compared to
the other modes.

\begin{figure}
  \centering\includegraphics[width=0.45\textwidth]{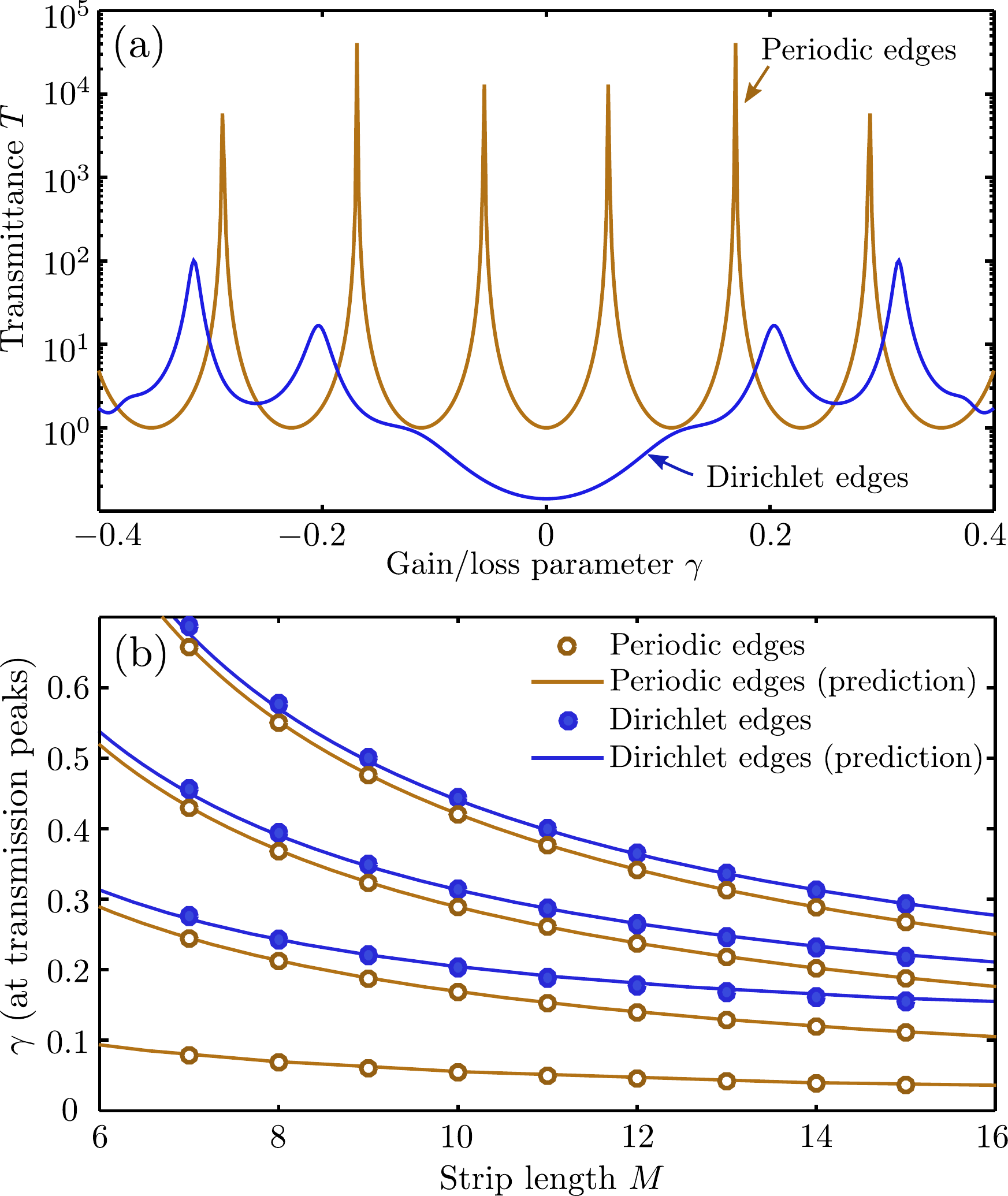}
  \caption{(a) Plot of the transmittance $T$ versus the gain/loss
    parameter $g$, for a strip of length $M=10$ and width $N=10$, and
    quasi-energy $\phi = -0.25\pi$.  Two cases are shown: periodic
    edges, and terminating (Dirichlet) edges.  (b) Plot of the values
    of $g$ at the transmittance peaks, versus the strip length $M$.
    The strip width is fixed at $N = 10$.  Open circles are for a
    strip with periodic edges, and closed symbols are for a strip with
    Dirichlet edges.  The curves show predictions made with the
    tachyonic dispersion relations, Eqs~(\ref{M1})--(\ref{M2}).}
  \label{fig:gvm}
\end{figure}

We can now formulate a transmission-based experimental signature for
the tachyon modes.  Consider a strip of length $M$ (in the $\hat{x}$
direction).  Along column 1, we inject equal wave amplitudes $a_1^n =
N^{-1/2}$ into each of the rightward links [see the schematic in
  Fig.~\ref{fig:projected}(a)]; then, $M$ columns to the right, we
calculate the total transmittance $T = \sum_{n=1}^N |a_1^n|^2$.
Physically, this corresponds to connecting the left and right edges of
a finite network to uniform multi-mode waveguides, which act as
scattering leads.  The total transmittance is then measured as a
function of the gain/loss parameter $\gamma$ (which is assumed to be
externally tunable, e.g.~by electrical or optical pumping).

The variation of the transmittance with $\gamma$ is shown in
Fig.~\ref{fig:gvm}(a).  Here, we take strip length $M = 10$ and width
$N = 10$, and set the network links to $\phi = -0.25\pi$,
corresponding to one of the Dirac points.  The transmission is found
to be peaked at certain values of $\gamma$; the peak positions depend
on the choice of periodic or Dirichlet edge conditions, as well as the
strip size.  Fig.~\ref{fig:gvm}(b) plots the values of $\gamma$ at the
transmission peaks, versus the strip length $M$.  For periodic edges,
the peaks can be fitted to
\begin{equation}
  k_c(\gamma) = \frac{(m+1/2)\pi}{M}, \quad m \in \mathbb{Z}_0^+,
  \label{M1}
\end{equation}
where $k_c(\gamma)$ is the tachyonic critical wavenumber given by
Eq.~(\ref{criticalk}).  For Dirichlet edges, on the other hand, the
transmission peaks can be fitted to
\begin{equation}
  k_c(\gamma)
  = \sqrt{\left(\frac{(m+1/2)\pi}{M}\right)^2 + \left(\frac{\pi}{N}\right)^2},
  \quad m \in \mathbb{Z}^+.
  \label{M2}
\end{equation}
The accuracy of these fits can be seen in Fig.~\ref{fig:gvm}(b), by
comparing the solid curves, which are produced from Eqs.~(\ref{M1})
and (\ref{M2}), to the circles, which correspond to the
numerically-obtained transmission peaks.

The relationship between the transmission peaks and the critical
wavenumber of the tachyon modes can be understood as follows.  At the
mid-gap quasi-energy $\phi = -0.25\pi$, only the tachyon modes are
propagating; the other modes are evanescent, and thus incapable of
forming standing-wave resonances.  The transmission peaks occur when
the strip length $M$ equals $(m+\tfrac{1}{2})/2$ tachyon mode
wavelengths, which allows for the largest intensity at the output
column (intensity anti-node) relative to the input column (intensity
node).  Although the tachyon modes have real wavenumbers, they do not
overlap exactly with the input amplitudes, so the presence of gain in
the network results in overall amplification, i.e.~transmission peaks
higher than unity.  For periodic edges, Eq.~(\ref{M1}) follows from
taking the tachyon modes to be plane waves propagating parallel to the
strip.  For Dirichlet boundary conditions, the tachyon modes must
undergo reflections from the strip edges, and taking the lowest-order
waveguide modes leads to Eq.~(\ref{M2}).  The dependence of this
equation on both the width $N$ and length $M$ emphasizes the fact that
the resonances arise from tachyon modes propagating in 2D, described
by a PT-symmetric 2D Dirac equation.

In conclusion, we have shown how an isolated tachyonic dispersion can
be realized in a photonic network, be it based on optical ring
resonators or microwave transmission lines.  A possible use of this
tachyonic dispersion may be delay lines of wide tunability with low
loss.

We are grateful to W.~Hu, H.~Wang, and Y.~Jia for helpful discussions.
This research was supported by the Singapore National Research
Foundation under grant No.~NRFF2012-02, and by the Singapore MOE
Academic Research Fund Tier 3 grant MOE2011-T3-1-005.


\begin{thebibliography}{99}
\bibitem{Bender} C.~M.~Bender and S.~Boettcher,
  Phys.~Rev.~Lett.~\textbf{80}, 5243 (1998).
\bibitem{Bender02} C.~M.~Bender, M.~V.~Berry, and A.~Mandilara,
  J.~Phys.~A \textbf{35}, L467 (2002).
\bibitem{ElGanainy} R.~El-Ganainy, K.~G.~Makris,
  D.~N.~Christodoulides, and Z.~H.~Musslimani, Opt. Lett. {\bf 32},
  2632--2634 (2007).
\bibitem{Kostas1} K.~G.~Makris, R.~El-Ganainy, D.~N.~Christodoulides,
  and Z.~H.~Musslimani, Phys.~Rev.~Lett.~{\bf 100}, 103904 (2008).
\bibitem{Musslimani} Z.~H.~Musslimani, K.~G.~Makris, R.~El-Ganainy,
  and D.~N.~Christodoulides, Phys.~Rev.~Lett.~\textbf{100}, 030402
  (2008); J.~Phys.~\textbf{A} 41, 244019 (2008).
\bibitem{Guo} A.~Guo, G.~J.~Salamo, D.~Duchesne, R.~Morandotti,
  M.~Volatier-Ravat, V.~Aimez, G.~A.~Siviloglou, and
  D.~N.~Christodoulides, Phys.~Rev.~Lett.~\textbf{103}, 093902 (2009).
\bibitem{Ruter} C.~E.~R\"{u}ter, K.~G.~Makris, R.~El-Ganainy,
  D.~N.~Christodoulides, M.~Segev, and D.~Kip, Nat.~Phys.~\textbf{6},
  192 (2010).
\bibitem{Kostas2} K.~G.~Makris \textit{et al.},
  Phys.~Rev.~Lett.~\textbf{100}, 103904 (2008);
  Phys.~Rev.~A~\textbf{81}, 063807 (2010).
\bibitem{Peschel} A.~Regensburger, C.~Bersch, M.-A.~Miri,
  G.~Onishchukov, D.~N.~Christodoulides, and U.~Peschel, Nature
  \textbf{488}, 167 (2012).
\bibitem{Peleg} Y.~Plotnik, \textit{et al.}, Nature Mat.~\textbf{13},
  57 (2014).
\bibitem{RechtsmanFTI} M.~C.~Rechtsman, J.~M.~Zeuner, Y.~Plotnik,
  Y.~Lumer, D.~Podolsky, F.~Dreisow, S.~Nolte, M.~Segev, and
  A.~Szameit,
  Nature {\bf 496}, 196 (2013).
\bibitem{ChalkerCo} J.~T.~Chalker, and P.~D.~Coddington,
  J.~Phys.~C {\bf 21}, 2665 (1988).
\bibitem{kramersurvey} B.~Kramer, T.~Ohtsukib, and S.~Kettemanna,
  Phys.~Rep.~{\bf 417}, 211 (2005).
\bibitem{HoChalker} C.-M.~Ho, and J.~T.~Chalker,
  Phys.~Rev.~B {\bf 54}, 8708
  (1996).
\bibitem{Pasek} M.~Pasek and Y.~D.~Chong,
  Phys.~Rev.~B {\bf 89},
  075113 (2014).
\bibitem{hafezi} M.~Hafezi, E.~A.~Demler, M.~D.~Lukin, and J.~M.~Taylor,
  Nat.~Phys.~{\bf 7}, 907 (2011).
\bibitem{Shayan} M.~L.~Cooper, G.~Gupta, M.~A.~Schneider,
  W.~M.~J.~Green, S.~Assefa, F.~Xia, Y.~A.~Vlasov, and S.~Mookherjea,
  Opt.~Ex.~\textbf{18}, 26505 (2010).
\bibitem{hafezi2} M.~Hafezi, S.~Mittal, J.~Fan, A.~Migdall, and
  J.~M.~Taylor,
  Nat.~Photonics {\bf 7}, 1001 (2013).
\bibitem{Mercedes} H.~Hodaei, M.-A.~Miri, M.~Heinrich,
  D.~N.~Christodoulides, and M.~Khajavikhan†,
  Science \textbf{21}, 975 (2014).
\bibitem{Zhang} L.~Feng, Z.~J.~Wong, R.-M.~Ma, Y.~Wang, and X.~Zhang,
  Science \textbf{21}, 972 (2014).
\bibitem{Liang} G.~Q.~Liang and Y.~D.~Chong,
  Phys.~Rev.~Lett.~{\bf 110}, 203904 (2013).
\bibitem{Liang2} G.~Q.~Liang and Y.~D.~Chong,
  Int.~J.~Mod.~Phys.~B {\bf 28}, 1441007 (2014).
\bibitem{Hu} W.~Hu, J.~C.~Pillay, K.~Wu, M.~Pasek, P.~P.~Shum, and
  Y.~D.~Chong,
  Phys.~Rev.~X {\bf 5}, 011012 (2015).
\bibitem{Gao} F.~Gao, \textit{et al.}, arXiv:1504.07809.
\bibitem{Ningyuan} N.~Jia, C.~Owens, A.~Sommer, D.~Schuster, and J.~Simon,
  Phys.~Rev.~X \textbf{5}, 021031 (2015).
\bibitem {Lindner} N.~H.~Lindner, G.~Refael and V.~Galitski,
  Nat.~Phys.~{\bf 7}, 490-495 (2011).
\bibitem{Cayssol} J.~Cayssol, B.~D\'ora, F.~Simon and R.~Moessner,
  Phys. Status Solidi RRL {\bf 7}, 101 (2013).
\bibitem{Levin} M.~S.~Rudner, N.~H.~Lindner, E.~Berg, and M.~Levin,
  Phys.~Rev.~X {\bf 3}, 031005 (2013).
\bibitem{haldane88} F.~D.~M.~Haldane, Phys. Rev. Lett. {\bf 61}, 2015 (1988).
\bibitem{tachyons} A.~Szameit, M.~C.~Rechtsman, O.~Bahat-Treidel, and
  M.~Segev,
  Phys.~Rev.~A \textbf{84}, 021806(R) (2011).
\bibitem{Zhen} B.~Zhen \textit{et al.}, arxiv:1504.00736.
\bibitem{superluminal} L.~J.~Wang, A.~Kuzmich and A.~Dogariu,
  Nature \textbf{406}, 277 (2000).
\bibitem{SchomerusGroupVelocity} H.~Schomerus and J.~Wiersig,
  Phys.~Rev.~A \textbf{90}, 053819 (2014).
\bibitem{Kostas3} K.~G.~Makris \textit{et al.}, in preparation.
\bibitem{PT0} H.~Schomerus, Phys.~Rev.~Lett.~\textbf{104}, 233601
  (2010).
\bibitem{PT1} Y.~D.~Chong, L.~Ge, and A.~D.~Stone,
  Phys.~Rev.~Lett.~\textbf{106}, 093902 (2011).
\bibitem{PT2} L.~Ge, Y.~D.~Chong, and A.~D.~Stone, Phys.~Rev.~A
  \textbf{85}, 023802 (2012).
\end{thebibliography}
\end{document}